\documentclass[aps,prl,twocolumn,showpacs,superscriptaddress]{revtex4-1}
\usepackage{bm}
\usepackage{mathrsfs}
\usepackage{amsmath}
\usepackage{amssymb}
\usepackage{graphicx}
\usepackage{amsfonts}
\usepackage{amsthm}
\usepackage{color}
\usepackage{dcolumn}
\usepackage{txfonts}
\usepackage{mathrsfs}

\begin{document}

\title{Universal Critical Behaviors in Non-Hermitian Phase Transitions}
\author{Bo-Bo Wei}
\email{Corresponding author: bbwei@szu.edu.cn}
\affiliation{School of Physics and Energy, Shenzhen University, Shenzhen 518060, China}
\author{Liang Jin}
\email{Corresponding author: jinliang@nankai.edu.cn}
\affiliation{School of Physics, Nankai University, Tianjin 300071, China}

\begin{abstract}
Quantum phase transitions also occur in non-Hermitian systems. In this work we show that density functional theory, for the first time, uncovers universal behaviors for phase transitions in non-Hermitian many-body systems. To be specific, we first prove that the non-degenerate steady state of a non-Hermitian quantum many-body system is a universal function of the first derivative of the steady state energy with respect to the control parameter. This finding has far-reaching consequences for non-Hermitian systems: (i) It bridges the nonanalytic behavior in physical observable and nonanalytic behavior of steady state energy, which explains why the quantum phase transitions in non-Hermitian systems occur for finite systems. (ii) It predicts universal scaling behaviors of any physical observable at non-Hermitian phase transition point with scaling exponent being $(1-1/p),2(1-1/p),,\cdots,n(1-1/p),\cdots$ with $p$ being the number of coalesced states at the exceptional point and $n$ being a positive integer. (iii). It reveals that quantum entanglement in non-Hermitian phase transition point presents universal scaling behaviors with critical exponents being $(1-1/p),2(1-1/p),\cdots,n(1-1/p),\cdots$. These results uncover universal critical behaviors in non-Hermitian phase transitions and provide profound connections between entanglement and phase transition in non-Hermitian quantum many-body physics and establish foundations for quantum metrology in non-Hermitian systems.
\end{abstract}


\maketitle

\emph{Introduction.}-Quantum phase transitions occurs when the ground state of a quantum many-body system experiences a sudden change as the parameter of the system is tuned through a critical point \cite{Sachedev2011}. It is one of the most intriguing phenomena in many-body physics because it indicates emergence of new states of quantum matter and new physics \cite{Sachedev2011,Wen2005}. In the study of quantum phase transitions, it is usually assumed that the Hamiltonian are Hermitian. However the non-Hermitian Hamiltonian indeed arises due to the spontaneous decay in current experimental results in cavities \cite{cavities1,cavities2}, waveguides \cite{waveguides1,waveguides2}, optomechanics \cite{mechanics1} and cold atoms \cite{coldatoms}. These experimental progresses provide new opportunity for discovering new classes of phase transitions beyond the Hermitian paradigm.

Non-Hermitian models draw a great deal of interest since they present richer behaviors \cite{Moiseyev2011,Berry2004,Bender2007,Heiss2012,richer1,richer2}, such as PT symmetry \cite{Bender1998,PT1,PT2}, localization\cite{local1}, dynamical phase transitions when the parameter are extended into the complex plane of physical parameters \cite{Wei2012,Heyl2013,Wei2014,Peng2015}. Recently, It was found that quantum phase transitions occurs in the steady state of non-Hermitian systems \cite{Tony1,Tony2}. However the universal critical behaviors of quantum phase transitions in the steady state of a general non-Hermitian systems have been illusive.

In this work we uncover the universal critical behavior of quantum phase transitions in the steady state of non-Hermitian many-body systems from density functional theory. We rigorously prove that the non-degenerate steady state of a non-Hermitian quantum man-body system is a universal function of the first derivative of the steady state energy with respect to the control parameter. Furthermore, we show that quantum entanglement in the non-degenerate steady state is also a universal function of first derivative of the steady state energy with respect to the control parameter. Because the non-Hermitian phase transition points are the exceptional point of the Hamiltonian \cite{Moiseyev2011,Berry2004,Bender2007,Heiss2012,richer1,richer2}, the first derivative of the steady state energy presents universal scaling behavior near the exceptional point \cite{Moiseyev2011,function}. Due to the universal dependence of the steady state on the first derivative of the steady state energy, we deduce the universal critical behaviors of physical observables and of quantum entanglement at non-Hermitian phase transitions point of the steady state.

\emph{Quantum Phase Transitions in Non-Hermitian Systems.}-
Let us consider a general non-Hermitian quantum many-body system with Hamiltonian,
\begin{eqnarray}
H(\gamma)=H_0+i\gamma H_1,
\end{eqnarray}
where $H_0$ and $H_1$ are Hermitian operators and $\gamma$ is a real control parameter. This non-Hermitian Hamiltonian can be realized as an effective Hamiltonian of an atomic systems without decay event \cite{Exp1,Exp2,Exp3,Exp4,Exp5}. Non-Hermitian Hamiltonian has eigenstates with complex eigenvalues. With time evolution, the weight in each eigenstate decreases over time because of the imaginary parts of the eigenvalues.
After a sufficient amount of time, the state consists mostly of the eigenstate whose eigenvalue has the largest imaginary part. This eigen state is termed the steady
state and denoted by $|\Psi_S\rangle$. We are interested in this
surviving eigenstate, because it is the one that would be
observed experimentally. Based on these concepts for non-Hermitian systems, we are ready to establish the first central theorem of this work.

\textbf{Theorem 1}: The non-degenerate steady state of a non-Hermitian quantum many-body system with Hamiltonian $H=H_0+i\gamma H_1$ is a universal function of the first derivative of the steady state energy with respect to the control parameter $\gamma$, i.e. $\frac{\partial E_S}{\partial\gamma}$.

In Theorem 1, the universal means that the function form of the dependence of steady state on the first derivative of the steady state energy does not change with variation of the control parameter so along as the steady state is in the same phase. The proof of Theorem 1 is based on the following two Lemmas.

\textbf{Lemma 1}: There is a one-to-one correspondence between the non-degenerate eigenket $|\Psi_S\rangle$ of the steady state in a non-Hermitian quantum many-body system with Hamiltonian $H=H_0+i\gamma H_1$ and the control parameter $\gamma$.\\
\emph{Proof:} For a given $\gamma$, by diagonalizing $H(\gamma)=H_0+i\gamma H_1$, we can get the steady state $|\Psi_S\rangle$; We also need to prove that the non-degenerate steady state also uniquely specifies the parameter $\gamma$. This is done by reductio and absurdum. We assume that two different parameters $\gamma$ and $\gamma'$ with $\gamma\neq\gamma'$ have the same steady state, $|\Psi_S\rangle$, then we have
\begin{eqnarray}
(H_0+i\gamma H_1)|\Psi_S\rangle&=&E(\gamma)|\Psi_S\rangle,\label{th1a}\\
(H_0+i\gamma' H_1)|\Psi_S\rangle&=&E(\gamma')|\Psi_S\rangle.\label{th1b}
\end{eqnarray}
Subtracting Equation \eqref{th1a} from Equation \eqref{th1b}, we get
\begin{eqnarray}
\Big[i(\gamma-\gamma')H_1-(E(\gamma)-E(\gamma'))\Big]|\Psi_S\rangle=0.
\end{eqnarray}
It means that $\gamma=\gamma'$ and $E(\gamma)=E(\gamma')$ and it contradicts the assumption. Thus Lemma 1 is proved. Since $\langle\widetilde{\Psi}_S|$ is the eigen bra of the steady state of $H(\gamma)$ with the maximum imaginary part, likewise, we can prove $\langle\widetilde{\Psi}_S|$ and $\gamma$ are also one-to-one mapped. These means that
\begin{eqnarray}
\gamma\Longleftrightarrow|\Psi_S\rangle,\\
\gamma\Longleftrightarrow\langle\widetilde{\Psi}_S|.
\end{eqnarray}

\textbf{Lemma 2}: There is a one-to-one map between the control parameter $\gamma$ and the density $\langle H_1\rangle_B=\langle\widetilde{\Psi}_S(\gamma)|H_1|\Psi_S(\gamma)\rangle$ in the non-degenerate steady state. \\
\emph{Proof}: For a given $\gamma$, $|\Psi_S\rangle$ and $\langle\widetilde{\Psi}_S|$ are uniquely specified according to Lemma 1. Then $\langle\widetilde{\Psi}_S|H_1|\Psi_S\rangle$ can be determined. We denote the eigen kets of $H$ at parameters $\gamma$ and $\gamma'$ by $|\Psi_S\rangle$ and $|\Psi_S'\rangle$ and eigen bras of $H$ at parameters $\gamma$ and $\gamma'$ by $\langle\widetilde{\Psi}_S|$ and $\langle\widetilde{\Psi}_S'|$. Now we have to show that if $\gamma\neq\gamma'$, $\langle H_1\rangle_{B}\neq\langle H_1\rangle_B'$. This can be done by reductio and absurdum. We assume two different control parameter $\gamma\neq\gamma'$ produce the same density
$\langle H_1\rangle_{B}=\langle H_1\rangle_B'$. According to maximum of the imaginary part of the steady state energy, we have
\begin{eqnarray}
\Im\langle\widetilde{\Psi}_S|H(\gamma)|\Psi_S\rangle&>&\Im\langle\widetilde{\Psi}_S'|H(\gamma)|\Psi_S'\rangle,\nonumber\\
&=&\Im\langle\widetilde{\Psi}_S'|H(\gamma')|\Psi_S'\rangle+\Im\Big[i(\gamma-\gamma')\langle\widetilde{\Psi}_S'|H_1|\Psi_S'\rangle\Big],\nonumber\\
&=&\Im E_S'+\Im\Big[i(\gamma-\gamma')\langle\widetilde{\Psi}_S'|H_1|\Psi_S'\rangle\Big].
\end{eqnarray}
Similarly by exchanging $\gamma$ and $\gamma'$ and their eigenstates, we get
\begin{eqnarray}
\Im\langle\widetilde{\Psi}_S'|H(\gamma')|\Psi_S'\rangle&>&\Im\langle\widetilde{\Psi}_S|H(\gamma')|\Psi_S\rangle,\nonumber\\
&=&\Im\langle\widetilde{\Psi}_S|H(\gamma)|\Psi_S\rangle+\Im\Big[i(\gamma'-\gamma)\langle\widetilde{\Psi}_S|H_1|\Psi_S\rangle\Big],\nonumber\\
&=&\Im E_S+\Im\Big[i(\gamma'-\gamma)\langle\widetilde{\Psi}_S|H_1|\Psi_S\rangle\Big].
\end{eqnarray}
Sum up the above two equations, we get
\begin{eqnarray}
\Im E_S+\Im E_S'>\Im E_S+\Im E_S'.
\end{eqnarray}
It is a contradiction and thus our assumption is wrong.  Lemma 2 is proved. $\gamma$ and $\langle H_1\rangle_B$ are one-to-one mapped.
\begin{eqnarray}
\gamma\Longleftrightarrow\langle H_1\rangle_B.
\end{eqnarray}

Now we are ready to prove Theorem 1.  Combining Lemma 1 and Lemma 2, we know that the steady state of a non-Hermitian quantum many-body system is uniquely specified by $\langle H_1\rangle_B$, namely
\begin{eqnarray}
|\Psi_S(\gamma)\rangle\Longrightarrow |\Psi_S(\langle H_1\rangle_B)\rangle,\\
\langle\widetilde{\Psi}_S(\gamma)|\Longrightarrow \langle\widetilde{\Psi}_S(\langle H_1\rangle_B)|.
\end{eqnarray}
Hellmann-Feynman Theorem for non-Hermitian system tells us for any eigenstate of $H(\gamma)$ \cite{Moiseyev2011},
\begin{eqnarray}\label{HF}
\Big\langle\widetilde{\Psi}_n\Big|\frac{\partial H(\gamma)}{\partial\gamma}\Big|\Psi_n\Big\rangle&=&\frac{\partial E_S(\gamma)}{\partial\gamma}.
\end{eqnarray}
Applying Equation \eqref{HF} to the steady state, we get
\begin{eqnarray}
\Longrightarrow\langle H_1\rangle_B=-i\frac{\partial E_S}{\partial\gamma}.
\end{eqnarray}
Thus the non-degenerate steady state is also uniquely specified by $-i\frac{\partial E_S}{\partial\gamma}$, i.e.
\begin{eqnarray}
|\Psi_S(\gamma)\rangle\Longrightarrow |\Psi_S(\langle H_1\rangle_B)\rangle\Longrightarrow \Big|\Psi_S\Big(-i\frac{\partial E_S}{\partial\gamma}\Big)\Big\rangle.
\end{eqnarray}
Theorem 1 is proved.

Theorem 1 is quite general and valid for any finite spin systems, Fermions or Bosons in lattices. Theorem 1 is in the same spirit as density functional theory developed by Honhenberg, Kohn and Sham \cite{HK1964,KS1965}. Here we prove that the one-to-one correspondence between the steady state and the density is also valid in non-Hermitian systems for the first time.

An immediate consequence of the Theorem 1 is that the steady state average value of any physical observable $O$ which does not commute with the Hamiltonian $[O,H]\neq0$ is also a universal function of the first derivative of the steady state energy with respect to the control parameter, $\frac{\partial E_S}{\partial\gamma}$,
\begin{eqnarray}
\langle O\rangle&=&\Big\langle\Psi_S\Big(\frac{\partial E_S}{\partial\gamma}\Big)\Big|O\Big|\Psi_S\Big(\frac{\partial E_S}{\partial\gamma})\Big\rangle.
\end{eqnarray}
This functional form is universal with respect to the control parameter $\gamma$ as along as the steady state is in the same phase and non-degenerate.

Non-Hermitian phase transition point, also called exceptional point, where two or more energy levels coalescence \cite{Moiseyev2011}. We assume there are $p\geq2$ levels coalescence at the exceptional point. Around the exceptional point, which is also an algebraic branch point, we can expand the steady state energy by \cite{Moiseyev2011,function},
\begin{eqnarray}
E_S(\gamma)&=&\sum_{i=0}^{\infty}\alpha_i(\gamma-\gamma_c)^{i/p},\\
&=&\alpha_0+\alpha_1(\gamma-\gamma_c)^{1/p}+\alpha_2(\gamma-\gamma_c)^{2/p}+\cdots.
\end{eqnarray}
If $\alpha_1\neq0$, we have
\begin{eqnarray}
\frac{\partial E_S}{\partial\gamma}\Big|_{\gamma\rightarrow\gamma_c}\propto (\gamma-\gamma_c)^{(1-p)/p}.
\end{eqnarray}
It diverges as $\gamma\rightarrow\gamma_c$. Since the average value of any physical observable is a universal function of the first derivative of the steady state energy, defining $Y\equiv\frac{\partial E_S}{\partial\gamma}$, then we have
\begin{eqnarray}
\langle O\rangle &=& f(Y).
\end{eqnarray}
Expanding $f(Y)$ around the critical point $Y\rightarrow\infty$, we thus get
\begin{eqnarray}\label{Ex1}
\langle O\rangle &=& f_0+f_1\frac{1}{Y}+f_2\frac{1}{Y^2}+\cdots,
\end{eqnarray}
where $f_0,f_1,f_2,\cdots$ are expansion coefficients and should be constant. So the steady state average of $O$ around the critical point is
\begin{eqnarray}
\delta\langle O\rangle&\propto&f_1(\gamma-\gamma_c)^{(p-1)/p}+f_2(\gamma-\gamma_c)^{2(p-1)/p}+\cdots,
\end{eqnarray}
where $\delta\langle O\rangle\equiv\langle O\rangle-\langle O\rangle_c$. Then the susceptibility of $\langle O\rangle$ is
\begin{eqnarray}
\chi=\frac{\partial\langle O\rangle}{\partial\gamma}\propto f_1(\gamma-\gamma_c)^{-1/p}+f_2(\gamma-\gamma_c)^{-2/p+1}+\cdots.
\end{eqnarray}
For different observables, the expansion coefficients in Equation \eqref{Ex1} are different. In particular, some of the expansion coefficients may vanish. Considering such a case, we thus have\\
\textbf{Corollary 1}: The steady state average of an arbitrary physical observable $\langle O\rangle$ at the non-Hermitian phase transition point presents scaling behavior, $\delta\langle O\rangle=\langle O\rangle-\langle O\rangle_c\propto(\gamma-\gamma_c)^{\alpha}$, with exponents $\alpha$ being $(1-1/p),2(1-1/p),\cdots,n(1-1/p),\cdots$ and $n$ being positive integers. \\
\textbf{Corollary 2}: The susceptibility of an arbitrary physical observable in the steady state at the non-Hermitian phase transition point scales as, $\delta\chi=\chi-\chi_c\propto(\gamma-\gamma_c)^{\beta}$, with exponents $\beta=-1/p,1-2/p,2-3/p,\cdots,(n-1)-n/p,\cdots.$ and $n$ being positive integers. \\
For $p=2$ case, there are two levels coalescence at the exceptional point and we then have
\begin{eqnarray}
\delta\langle O\rangle\propto(\gamma-\gamma_c)^{\alpha},
\end{eqnarray}
where $\alpha=\frac{1}{2},1,\frac{3}{2},2,\cdots$
and the susceptibility near the non-Hermitian phase transition point scales as,
\begin{eqnarray}
\chi=\frac{\partial\langle O\rangle}{\partial\gamma}\propto(\gamma-\gamma_c)^{\beta},
\end{eqnarray}
where $\beta=-\frac{1}{2},0,\frac{1}{2},1,\cdots.$
This means that the first derivative of an arbitrary physical quantity diverges at a behavior $(\gamma-\gamma_c)^{-1/2}$ in non-Hermitian phase transition point. This reveals how the non-Hermitian coalescence in a finite system leads to the non-analytic behavior of physical observable, thus non-Hermitian phase transitions.

\emph{Quantum entanglement in Non-Hermitian Systems.}-Quantum entanglement provides a powerful way to understand the nature of many-body systems. In particular, it has been shown that
entanglement are deeply related to phase transitions in condensed matter systems \cite{Vedral2008}. Recently it was also found that the entanglement in non-Hermitian phase transitions is bigger than that of Hermitian quantum phase transitions \cite{Tony2}. We first establish a theorem which connects the entanglement and quantum phase transitions in non-Hermitian systems.\\
\textbf{Theorem 2}: Any entanglement measure in the non-degenerate steady state of a non-Hermite quantum many-body system with Hamiltonian $H(\gamma)=H_0+i\gamma H_1$ is a universal function of first derivative of steady state energy with respect to the control parameter,
\begin{eqnarray}
M=M\Big(\frac{\partial E_s}{\partial\gamma}\Big).
\end{eqnarray}
\emph{Proof:} The proof follows from the fact that, according to Theorem 1, the steady state $|\Psi_S\rangle$ in non-Hermitian systems is a unique function of $\frac{\partial E_S}{\partial \gamma}$ and also $|\Psi_S\rangle$ provides the complete information of the system in the steady state, everything else is a unique function of $\frac{\partial E_S}{\partial \gamma}$. Formally let us consider an $n$-partite entanglement in spin-1/2 systems. For other cases, the proof can be generalized immediately. First of all any entanglement measure of $n$ qubits is always a function of the matrix elements of the reduced density matrix of these qubits, $\rho_{12\cdots n}$: $M(\rho_{12\cdots n})$. For spin-1/2 systems, the $n$-body reduced density matrix can be written as
\begin{eqnarray}
\rho_{12\cdots n}=\sum_{a_1,a_2,\cdots,a_n=0,x,y,z}C_{a_1a_2,\cdots,a_n}\sigma_1^{a_1}\sigma_2^{a_2}\cdots\sigma_{n}^{a_n},
\end{eqnarray}
with
\begin{eqnarray}
C_{a_1a_2,\cdots,a_n}&=&\text{Tr}_{12\cdots n}[\rho_{12\cdots n}\sigma_1^{a_1}\sigma_2^{a_2}\cdots\sigma_n^{a_n}],\\
&=&\text{Tr}[\rho_S\sigma_1^{a_1}\sigma_2^{a_2}\cdots\sigma_n^{a_n}],\\
&=&\langle\sigma_1^{a_1}\sigma_2^{a_2}\cdots\sigma_n^{a_n}\rangle,
\end{eqnarray}
where $a_1,a_2,\cdots,a_n=0,x,y,z$ with $\sigma^0=I$ and $\rho_S=|\Psi_S\rangle\langle\Psi_S|$. Thus $M=M\Big(\langle\sigma_1^{a_1}\sigma_2^{a_2}\cdots\sigma_n^{a_n}\rangle\Big)$. According to Theorem 1, the average value of any observable can be taken as a function of $\frac{\partial E_S}{\partial \gamma}$. Therefore $M=M\Big(\frac{\partial E_S}{\partial \gamma}\Big)$. Theorem 2 is proved. Relations between entanglement and quantum phase transitions in Hermitian models are established in \cite{Wu2005} and was generalized to finite temperatures in \cite{Wei2016}.

Since entanglement for a physical state can only be finite and near the non-Hermitian phase transition point $Y=\frac{\partial E_S}{\partial\gamma}\propto(\gamma-\gamma_c)^{(1-p)/p}$ diverges, then we can expand the entanglement measure around the non-Hermitian phase transition point by
\begin{eqnarray}\label{Ex2}
M(Y)=m_0+\frac{m_1}{Y}+\frac{m_2}{Y^2}+\cdots,
\end{eqnarray}
where $m_0,m_1,\cdots$ are the expansion coefficients and should be constant. So the entanglement around the non-Hermitian phase transition point scales with the control parameter as
\begin{eqnarray}
\delta M\propto m_1(\gamma-\gamma_c)^{(p-1)/p}+m_2(\gamma-\gamma_c)^{2(p-1)/p}+\cdots,
\end{eqnarray}
wher $\delta M\equiv M(\gamma)-M(\gamma_c)$ and the first derivative of the entanglement measure scales as
\begin{eqnarray}
\frac{\partial M}{\partial\gamma}\propto m_1(\gamma-\gamma_c)^{-1/p}+m_2(\gamma-\gamma_c)^{1-2/p}+\cdots.
\end{eqnarray}
The expansion coefficients in Equation \eqref{Ex2} are different for different entanglement measures. In particular, some of the expansion coefficients may vanish. Considering such a case, we thus have\\
\textbf{Corollary 3}: Any entanglement measure of the steady state near the non-Hermitian phase transition point scales as, $\delta M=M(\gamma)-M(\gamma_c)\propto(\gamma-\gamma_c)^{\mu}$, with exponents $\mu=(1-1/p),2(1-1/p),3(1-1/p),\cdots,n(1-1/p),\cdots$ and $n$ being a positive integer. \\
\textbf{Corollary 4}: The first derivative of any entanglement measure of the steady state near the non-Hermitian phase transition point scales as, $\frac{\partial M}{\partial\gamma}-\frac{\partial M}{\partial\gamma}\Big|_{\gamma=\gamma_c}\propto(\gamma-\gamma_c)^{\nu}$, with exponents $\nu=-1/p,1-2/p,2-3/p,(n-1)+n/p,\cdots$ and $n$ being a positive integer.

Theorem 2 and corollary 3 and 4 establish rigourously the connections between quantum entanglement and quantum phase transition in non-Hermmitian systems. They are valid for any finite spin systems and Fermions or Bosons in a lattices.

\begin{figure}
\begin{center}
\includegraphics[scale=0.22]{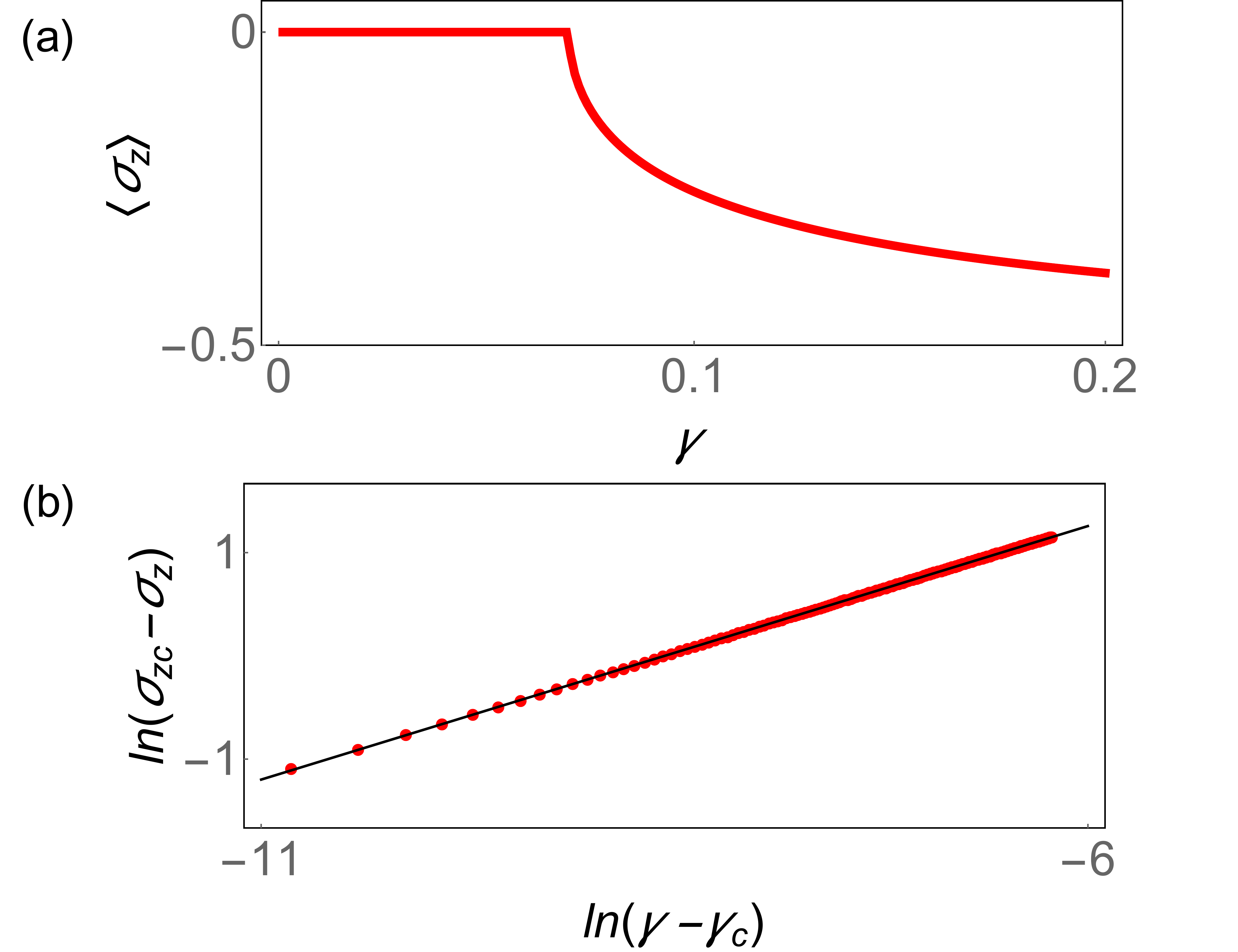}
\end{center}
\caption{(color online). Phase transitions in a Non-Hermitian LMG model. (a). The average magnetization along $z$ axis $\langle\sigma_z\rangle\equiv\langle J_z\rangle/N$ as a function of $\gamma$ in the LMG model with $N=40$ spins. (b). Scaling of the magnetization around the critical point $\gamma_c$. The vertical axes is the logarithmic of the difference between the magnetization at critical point and near the critical point. The horizontal axis is $\ln(\gamma_c-\gamma)$ where $\gamma_c$ is the critical control parameter. The red solid circle presents the numerical exact solution and the black solid line is the linear fitting line, where the slope is $0.4907637$. }
\label{fig:epsart1}
\end{figure}

\emph{Model study}-To demonstrate the above ideas, we study the LMG model with the Hamiltonian \cite{LMGmodel,Tony2}
\begin{eqnarray}\label{LMG}
H=\frac{V}{N}(J_x^2-J_y^2)-\frac{i\Gamma}{2}J_z-\frac{i\Gamma N}{4},
\end{eqnarray}
where $V$ is the coupling strength and $J_{\alpha}\equiv\frac{1}{2}\sum_{i=1}^N\sigma_i^{\alpha}, \alpha=x,y,z$ are the collective spin operators.
We consider $V$ as fixed and $\Gamma$ as varying parameter. In terms of raising and lowering operators of the collective spin, $J_{\pm}\equiv J_x\pm J_y$, we have
\begin{eqnarray}
H/V=\frac{1}{4N}(J_+^2+J_-^2)-\frac{i\gamma}{2}J_z-\frac{i\gamma N}{4}.
\end{eqnarray}
Here $\gamma=\Gamma/V$ being dimensionless control parameter. For convenience, we focus on the Dicke manifold with maximum angular momentum, so the
Hilbert space has dimension $N+1$.

Figure 1 shows the steady state average value of $\langle \sigma_z\rangle=\langle J_z\rangle/N$ in the LMG model with $N=40$ spins as a function of the control parameter $\gamma$. One can see that there is a critical $\gamma_c$. If $\gamma<\gamma_c$, $\langle\sigma_z\rangle=0$ and being smaller than zero if $\gamma>\gamma_c$. In Figure 1(b), we study the critical exponents of $\langle\sigma_z\rangle$ and plot $\ln(\langle\sigma_{z}\rangle_c-\langle\sigma_z\rangle)$ as a function of $\ln(\gamma-\gamma_c)$ near the critical point. We made a linear fit and found that the critical exponents being $0.4907637$. And it indicates near the critical point $\langle\sigma_z\rangle\propto(\gamma-\gamma_c)^{1/2}$. This is consistent with the prediction from Corollary 1 since there are two levels coalescence at the critical point \cite{Tony2}.

To quantify many-body entanglement, we study the averaged quantum Fisher information which is defined by \cite{QF1,QF2},
\begin{eqnarray}
F=\frac{4}{3N^2}[(\Delta J_x)^2+(\Delta J_y)^2+(\Delta J_z)^2],
\end{eqnarray}
where $N$ is the number of spins. In Figure 2, we present the quantum Fisher information of the steady state in the non-Hermitian LMG model with $N=40$ spins as a function of the control parameter. One can see that the quantum Fisher information is maximum when $\gamma<\gamma_c$ and decreases when $\gamma>\gamma_c$. In Figure 2(b), we study how the quantum Fisher information scales near the critical point where the quantum Fisher information is maximum and denoted by $F_C$. We plot $\ln(F_C-F)$ as a function of $\ln(\gamma-\gamma_c)$ near the critical point. We made a linear fit and found that the critical exponents being $0.981527$. And it indicates near the critical point $F_C-F\propto(\gamma-\gamma_c)^1$. This is consistent with the prediction from Corollary 3 since there are two levels coalescence at the critical point \cite{Tony2}.

\begin{figure}
\begin{center}
\includegraphics[scale=0.25]{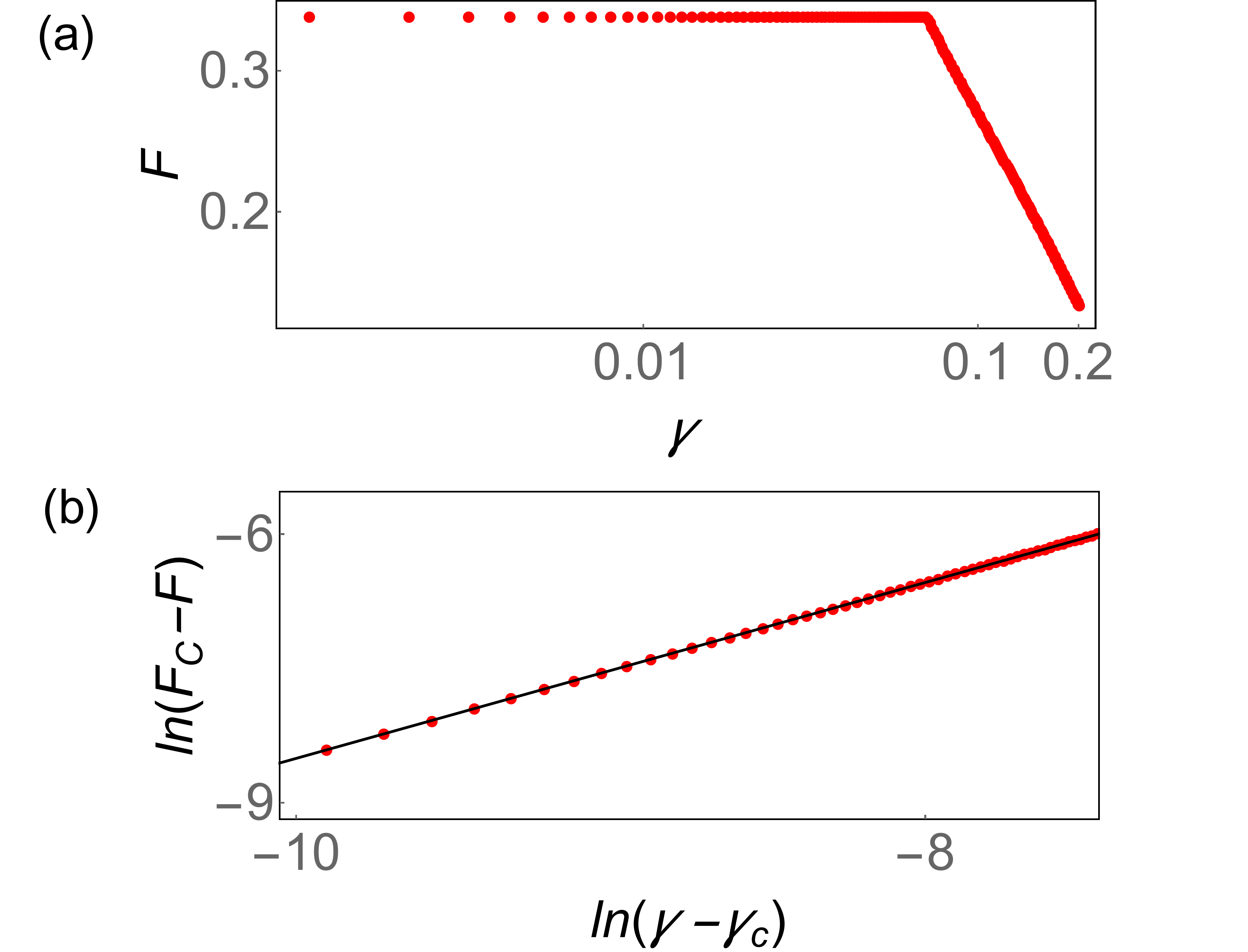}
\end{center}
\caption{(color online). Multipartite entanglement in non-Hermitian phase transitions. (a). Quantum Fisher information $F_Q$ as a function of the control parameter $\gamma$ in the non-Hermitian LMG model for $N=40$ spins.  (b). Scaling of quantum Fisher information near the non-Hermitian phase transition point. The vertical axes is the logarithmic of the difference between quantum Fisher information at critical point and the quantum Fisher information near the critical point and the horizontal axes is $\ln(\gamma_c-\gamma)$ where $\gamma_c$ is the critical control parameter. The red solid circle presents the numerical exact solution and the black solid line is the linear fitting line, where the slope is $0.981527$. }
\label{fig:epsart1}
\end{figure}

\emph{Conclusions}.-In this work we uncover universal critical behaviors for quantum phase transitions in non-Hermitian many-body systems. We prove that the non-degenerate steady state of an non-Hermitian quantum many-body system is a universal function of the first derivative of the steady state energy with respect to the control parameter. This finding bridges the nonanalytic behavior in physical observable with nonanalytic behavior of steady state energy and explains why the quantum phase transitions in non-Hermitian systems occurs in finite systems and predicts a universal scaling behavior of any physical observable and quantum entanglement near the non-Hermitian phase transition point. These results provide profound connections between entanglement and phase transition in non-Hermitian quantum many-body physics and establishes foundations for quantum metrology in non-Hermitian systems.

\begin{acknowledgements}
B.B.W. was supported by National Natural Science Foundation of China (Grants No. 11604220) and the Startup Funds of Shenzhen University. L. J. was supported by National Natural Science Foundation of China (Grants No. 11605094).
\end{acknowledgements}

\end{document}